\documentclass[aps,prb,preprint,groupedaddress]{revtex4-1}
\usepackage{braket}
\usepackage{amsmath}
\usepackage{graphicx}
\usepackage{color}

\begin{document}

\preprint{}

\title{Origin of layer dependence in band structures of two-dimensional materials}


\author{Mit H. Naik and Manish Jain}
\affiliation{Center for Condensed Matter Theory, Department of Physics, Indian Institute of Science, Bangalore 560012}


\date{\today}

\begin{abstract}
We study the origin of layer dependence in band structures of two-dimensional
materials. We find that the layer dependence, at the density functional theory
(DFT) level, is a result of quantum confinement and the non-linearity of the
exchange-correlation functional. 
We use this to develop an efficient scheme for
performing DFT and GW calculations of multilayer systems. We show that the 
DFT and quasiparticle band structures of a multilayer system can be derived from 
a single calculation on a monolayer of the material. 
We test this scheme on multilayers of MoS$_2$, graphene and phosphorene. 
This new scheme yields results 
in excellent agreement with the standard methods at a fraction of the computation cost. 
This helps overcome the challenge of performing fully 
converged GW calculations on multilayers of 2D materials, particularly in the 
case of transition metal dichalcogenides which involve very stringent convergence 
parameters. 
\end{abstract}


\maketitle

\section{Introduction}
Two-dimensional (2D) materials have been extensively studied in the last decade
\cite{PNAS.Novoselov,PRL.Mak,NL.Splendani,NN.Wang,PRL.Jin,
PRB.Tran,PRB.Kuc,PRL.Qiu,PRB.Tran,ACSNano.Liu} owing to their applications 
in electronics and optoelectronics \cite{NN.Radisavljevic,NN.Li,NN.Roy}.  
2D materials consist of layers that are held together by weak van der Waals 
forces.  A remarkable feature of these layered materials is the difference 
in properties of a monolayer compared to multilayers of the same material 
\cite{PRL.Mak,NL.Splendani,NN.Wang,PRL.Jin,APL.Ellis,
PRB.Tran,PRB.Kuc,PRB.Hong}. For instance,
monolayer of MoS$_2$ has a direct band gap, while multilayers of MoS$_2$ have
an indirect gap \cite{PRL.Jin,PRL.Mak,NL.Splendani,APL.Ellis,
PRB.Padilha}.  Most gapped 2D materials, like transition metal dichalcogenides (TMDCs), hexagonal boron nitride 
and phosphorene, show an unmistakable reduction
in band gap with the number of layers \cite{PRB.Padilha,PRB.Tran,2D.Tran,PRB.Kuc,PRB.Hong,NL.Splendani,PRL.Mak,APL.Ellis,PRB.Komsa,JPCL.Kang,JPCL.Fan,PRL.Gabriel}. 

First principles electronic structure calculations, based on the GW \cite{hedin65, 
*hedin70, PRB.Hybertsen,*PRL.Hybertsen}
approximation, have resulted in band gaps that are in excellent agreement with
experiments \cite{PRL.Qiu,PRB.Tran,2D.Tran,
NL.Aaron,NMat.Ugeda,PRL.Lischner,Nature.Ye,NL.Cao} on these materials. Band gaps 
of these materials calculated using density functional theory (DFT) 
\cite{PR.Hohenberg,PRB.Kohn}, while underestimated, also show a clear 
reduction with the number of layers \cite{PRB.Padilha,JPCL.Kang,JPCL.Fan,
PRB.Kuc,PRB.Hong,PRL.Gabriel}. Most studies have
attributed this reduction in the band gap to quantum confinement
\cite{PRL.Jin,PRB.Tran,PRB.Kuc,NL.Splendani}.  However, 
there are no studies that {\em quantitatively explain} this trend.

The idea of quantum confinement comes from the model of an electron in a
one-dimensional box. In multilayer stacks
of 2D materials, the confining length of the box is directly proportional to
the number of layers. In this context, the increase in the
confinement length in multilayers as compared to a monolayer, due the summing
of constituent layer potentials, is attributed to quantum confinement.
On adding the potentials of constituent layers, the inter-layer spacing creates
a finite barrier in the interstitial region between adjacent layers. This barrier with a
DFT calculated Hartree potential profile has been used to {\em qualitatively}
explain the trend in the layer dependence of the band gap in phosphorene
\cite{PRB.Tran}.  Recently, perturbation approaches 
\cite{JMR.Georgios,JPCL.Kang}, which need input from explicit multilayer calculations,
have been used to study layer dependence of band structures. 

Furthermore, in the case of TMDCs, it has been shown that the quasiparticle
band gap calculated using the GW approximation converges extremely slowly with
the number of unoccupied states, k-point sampling and the screened Coulomb
cut-off \cite{PRL.Erratum.Qiu,PRB.Qiu}. Performing separate GW calculations for
a monolayer, bilayer, trilayer or more with parameters that ensure convergence
is computationally very challenging. To study the variation of properties as a 
function of interlayer spacing or stacking of the layers needs one to 
repeat the calculation for different parameters.
Moreover, GW calculations on heterostructures 
of 2D materials are presently intractable due to their lattice incommensurability 
necessitating the use of large supercells of each material. Identifying the origin of layer
dependence opens up the possiblity of deriving ground state and excited state
properties of multilayer stacks from calculations on a monolayer alone. Doing
so would immensely ease the computation cost incurred in performing separate
DFT and GW calculations for different configurations of the constituent layers. 

We study the physical origin of layer dependence in band structures of
2D materials.  We show that while quantum confinement gives {\em qualitatively} the
right trend, the non-linearity of the $V_{xc}$ functional plays a
crucial role in {\em quantitatively} determining the layer dependence.
We show that within DFT, band structures of multilayer stacks can be derived from a
single calculation on a monolayer of the material. We also extend this scheme
to obtain quasiparticle energies of multilayer systems from a single GW
calculation on a monolayer of the material. We apply this method to understand
the layer dependence of band structure in multilayers of a prototypical
TMDC, 2H-MoS$_2$, and compare the results to the
standard calculations on this material \cite{PRL.Mak,NL.Splendani,NN.Wang}. We 
demonstrate the transferability of this scheme by applying it to graphene and 
phosphorene.

\section{Computation details}
We performed the DFT calculations using the plane-wave pseudopotential
method as implemented in Quantum Espresso \cite{QE.Giannozi} package. 
Norm-conserving pseudpotentials were used in all calculations. The wave
functions were expanded in plane-waves with an energy cut-off of 150 Ry for MoS$_2$. 
For graphene and phosphorene, we used a wavefunction cut-off of 70 Ry and 40 Ry 
respectively. The local density approximation to the exchange-correlation
functional \cite{PRB.Perdew} was used in MoS$_2$ and graphene calculations. 
For phosphorene we used 
the Perdew-Burke-Ernzerhof exchange-correlation functional \cite{JCP.PBE}. 
The Brillouin zone was sampled with 
$24\times24\times1$, $21\times21\times1$ and $21\times15\times1$ k-point 
grids for MoS$_2$, graphene and phosphorene respectively. We kept the in-plane 
lattice parameter for each material fixed in all the calculations. The bilayer 
and trilayer were constructed from the monolayer with the appropriate stacking
and inter-layer spacing. No atomic relaxations were allowed in the bilayer and 
trilayer. The supercell dimension in the out-of-plane direction 
was fixed at 35\AA.

The GW calculations were performed using the BerkeleyGW package
\cite{BGW.Deslippe,PRL.Samsonidze}. For MoS$_2$, the dielectric function was 
evaluated with plane waves upto a cutoff of 35 Ry and was extended to 
finite frequencies using the generalized plasmon pole (GPP) \cite{PRB.Hybertsen, PRL.Hybertsen} model. 
The self-energy was constructed using the one shot $\mathrm{G}_0 \mathrm{W}_0$ method. 
The Coulomb interaction was truncated in the out-of-plane direction \cite{PRB.Ismail-Beigi}. 
For MoS$_2$ supercell size of 35~\AA~in the out-of-plane direction, $24\times24\times1$ 
k-point sampling and 8400 valence and conduction states are necessary to ensure
convergence \cite{PRB.Qiu}. We perform separate calculations on monolayer, bilayer
and trilayer MoS$_2$ with $12\times12\times1$ k-point sampling
and 6000 valence and conduction states and compare the quasiparticle band
structures of bilayer and trilayer with results obtained from the proposed scheme. These
parameters, while not fully converged, demonstrate the efficacy of this scheme.
We also perform fully converged calculations on a monolayer of MoS$_2$ and derive the 
band gap, ionization potential and electron affinity of bilayer and trilayer 
using our scheme. Our fully converged monolayer band gap is in good agreement 
with previous calculations \cite{PRL.Qiu,PRB.Qiu}. For phosphorene, we
perform separate calculations on monolayer, bilayer and trilayer with 
$21\times15\times1$ k-point sampling, 800 valence and conduction states, and 
15 Ry dielectric cutoff. We compare the band gap, ionization potential and 
electron affinity obtained using our scheme to those obtained from the full 
calculation. The static remainder technique was used to speed up convergence with the 
number of bands \cite{PRB.Deslippe} in all calculations. 

\section{DFT band structures}

\begin{figure}
 \centering
 \includegraphics[scale=0.23]{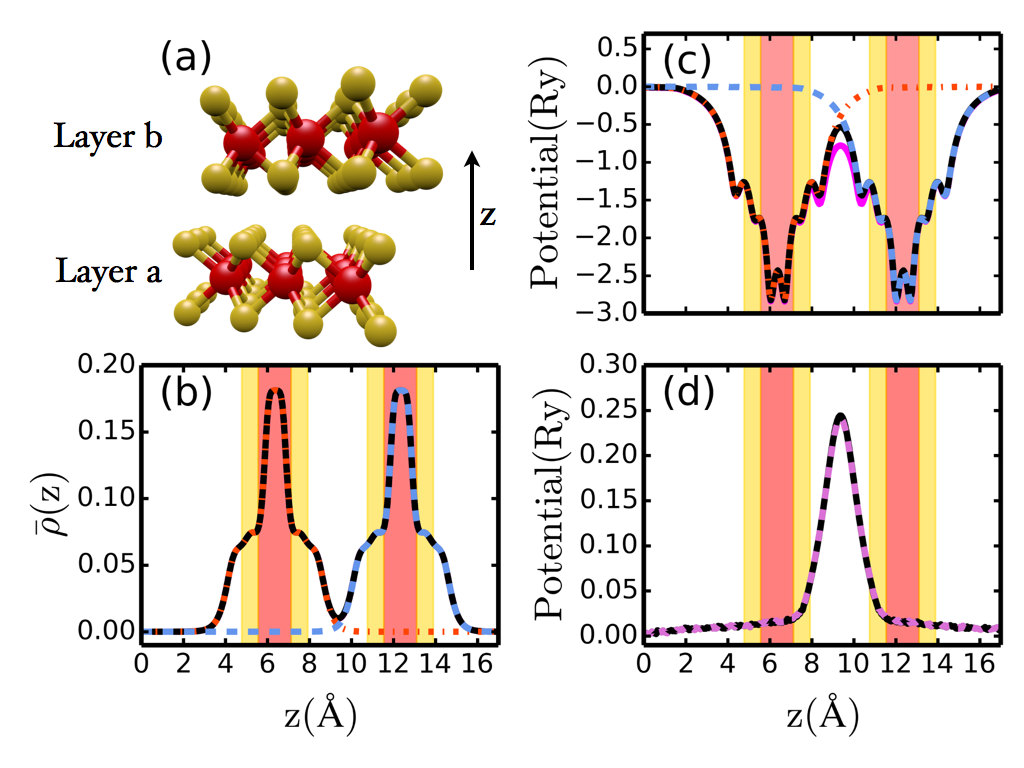}
 \caption {\label{fig1} (Color online) (a) Bilayer MoS$_2$ in AB stacking, red spheres
 represent Mo atoms and yellow spheres S atoms. (b) Planar averaged charge
 density of bilayer MoS$_2$ (black solid line), planar averaged charge density
 of layer 'a' (red dash-dot line), layer 'b' (blue dashed line) and sum of layer
 'a' and 'b' (magenta line not seen). The shaded regions indicate the position
 of the MoS$_2$ layers in the simulation cell.  (c) Planar averaged DFT
 potential of the layer 'a', $V_{tot}^a$ (red dash-dot line), layer 'b',
 $V_{tot}^b$ (blue dashed line), the sum $V_{tot}^a+V_{tot}^b$ (solid magenta
 line) and bilayer, $V_{tot}^{bi}$ (black solid line). (d) Difference
 $V_{tot}^{bi} - (V_{tot}^a + V_{tot}^b)$ (black solid line) and $
 V_{xc}[\rho^a + \rho^b] - (V_{xc}[\rho^a]+V_{xc}[\rho^b])$ (violet dashed
 line).} 
\end{figure}

In order to understand the layer dependence of the DFT band gap, consider a
bilayer of MoS$_2$ with constituent layers labelled 'a' and 'b' (Fig.
\ref{fig1}(a)) \cite{PRB.He}.  We perform separate DFT calculations on layer
'a', layer 'b' and the bilayer, to obtain the self-consistent charge densities
and potentials of each. In Fig. \ref{fig1}(b), we plot the planar averaged
charge densities of layer 'a', layer 'b', the bilayer. From the figure one can
see that the bilayer charge density, $\rho^{bi}$, lies on top of the
sum of the charge densities of the constituent layers, $\rho^{a}$+$\rho^{b}$,
which indicates that there is no significant rearragement of charge in the
bilayer compared to the monolayers. 
 Fig.  \ref{fig1}(c) shows the planar
averaged total DFT potential of the bilayer, $V_{tot}^{bi}$. From this figure,
it is clear that the sum of the layer potentials, $V_{tot}^a + V_{tot}^b$, is
not equal to the bilayer potential. The sum of the potentials, as described above, 
is the use of quantum confinement to obtain the potential for the bilayer.
The difference, $V_{tot}^{bi} - (V_{tot}^a
+ V_{tot}^b)$, is localized to the interstitial region between the two layers 
where the charge densities overlap. It can not arise from the 
Hartree potential since it is a linear functional of the charge density and 
here the charge density of the bilayer is a sum of the charge densities of 
the individual layers. It can not arise from the ionic potential either, due to 
the absence of atomic relaxations in the bilayer. The difference must 
arise due to the non-linearity of the exchange-correlation functional.
Fig. \ref{fig1}(d) plots this difference which is an additional 
barrier between the layers. Fig. \ref{fig1}(d) also plots 
$ V_{xc}[\rho^a + \rho^b] - (V_{xc}[\rho^a]+V_{xc}[\rho^b]) = \Delta V_{xc}$, 
showing that this additional
barrier indeed comes solely from the exchange-correlation difference. The
total bilayer potential can thus be expressed as a sum of the individual layer
potentials and $\Delta V_{xc}$. Thus, the layer dependence of properties 
is an effect of quantum
confinement and non-linearity of the $V_{xc}$ functional. 
\begin{figure}
  \centering
 \includegraphics[scale=0.24]{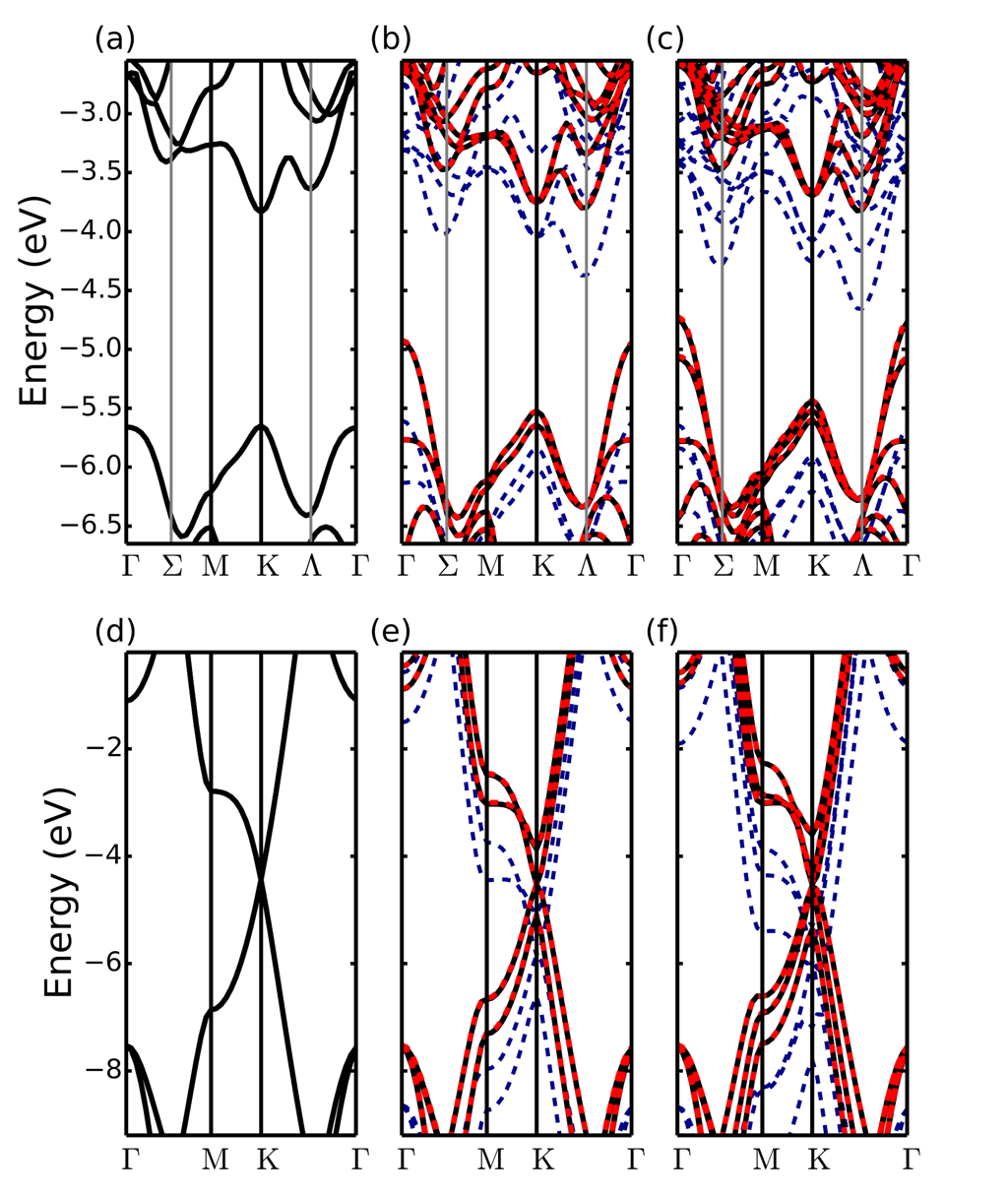}
 \caption{\label{fig2} (Color online) Panels (a), (b) and (c) plot the DFT
 band structure of monolayer, bilayer and trilayer MoS$_2$ respectively. 
 Panels (d), (e) and (f) plot the DFT band structures of monolayer, bilayer and
 trilayer graphene. The
 black solid line indicates the result from separate DFT calculations on these
 systems. The red dashed line shows the result obtained by a single-shot
 diagonalisation of the constructed Hamiltonian Eq \ref{eqn1} for these systems in the
 basis of the wavefunctions of the constituent layers. The blue dashed lines
 are the eigenvalues obtained by considering only quantum confinement in these
 systems. } 
\end{figure}

We can construct the DFT Hamiltonian for the bilayer in terms of the potential
and charge density of the constituent layers as: 
\begin{equation}
 \label{eqn1}
 H = \widehat{T} + V_{tot}^a + V_{tot}^b + \Delta V_{xc}
\end{equation}
where $\widehat{T}$ is the kinetic energy operator. 
It should be noted that
everything required to construct this Hamiltonian can be obtained from a single
monolayer calculation, say on layer 'a'. Based on the relative configuration of
atoms in layer 'b' with respect to layer 'a', a suitable transformation can be
applied on $\rho^a$ and $V_{tot}^a$ to obtain $\rho^b$ and $V_{tot}^b$
respectively.  The wavefunctions of layer 'a', \{$\psi^a_{nk}$\}, can similarly
be transformed to obtain the wavefunctions of layer 'b', \{$\psi^b_{nk}$\}. The
Hamiltonian can then be constructed in the basis of the wavefunctions of the
two layers, \{$\psi^a,\psi^b$\}, keeping in mind that the wavefunctions of the
two layers do not form an orthogonal basis. The generalized eigenvalue problem
can be solved to yield eigenvalues and eigenfunctions of the bilayer.  This
procedure can easily be generalized to N layers: $H = \widehat{T} + \sum\limits_{i=1}
^{N}V_{tot}^i + \Delta V_{xc}$; where $\Delta V_{xc} = V_{xc}[\sum\limits_{i=1}^{N}\rho^i]
- \sum\limits_{i=1}^{N}V_{xc}[\rho^i]$. It is worth noting that while constructing the 
Hamiltonian we use $V_{xc} (\mathbf{r})$ and not just the planar averaged quantities.
\begin{figure}
  \centering
  \includegraphics[scale=0.30]{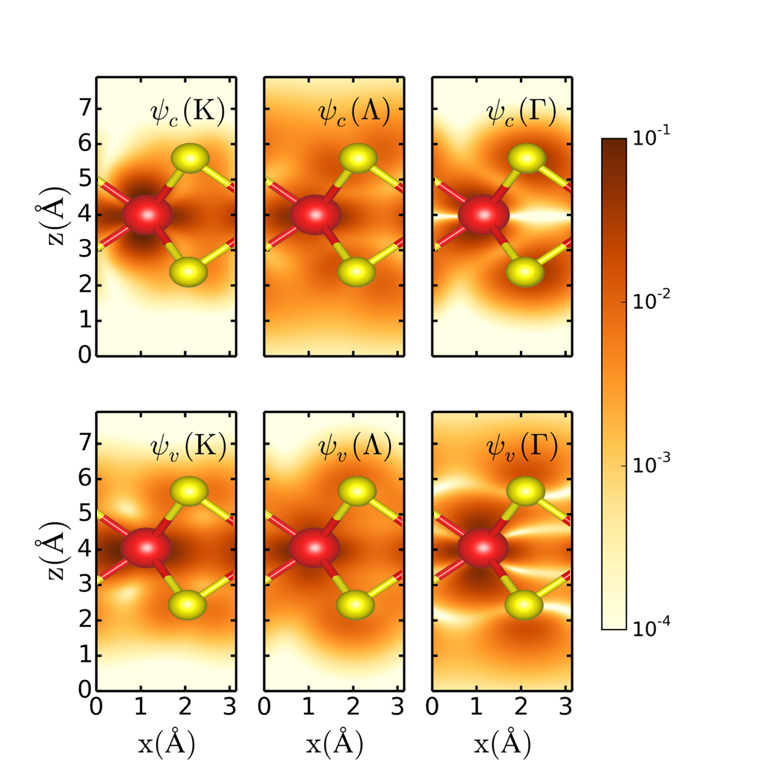}
  \caption{\label{fig3} (Color online) Modulus squared wavefunctions of
monolayer MoS$_2$, integrated out along [010] lattice vector direction. z is
the out-of-plane direction and x the [100] direction. Top panel: CBM
wavefunctions at K, $\Lambda$ and $\Gamma$ points, respectively. Bottom panel:
VBM wavefunctions at K, $\Lambda$ and $\Gamma$ points, respectively. }
\end{figure}
  
The band structure of monolayer, bilayer and trilayer MoS$_2$ from separate DFT
calculations is plotted in Fig. \ref{fig2} (a), (b) and (c) respectively. The
points $\Sigma$ and $\Lambda$ are marked halfway between $\Gamma$-M and
K-$\Gamma$ respectively.  Fig. \ref{fig2} (d), (e) and (f) similarly show the band 
structure of monolayer, bilayer and trilayer graphene respectively. Fig. \ref{fig2} also 
shows the band structure obtained by neglecting $\Delta V_{xc}$ from
 Eqn. (\ref{eqn1}). This describes the effect of quantum confinement 
alone on the layer dependence of the band
structures. While it captures the qualitative trends like the transition from
direct to indirect band gap in MoS$_2$, it fails to give an accurate layer
dependence of the value of the band gap, ionization potentials and level
splittings in the band structure. In MoS$_2$, the splittings are overestimated 
at the conduction band  minimum (CBM) of the $\Lambda$ point and 
underestimated at the $\Gamma$ point valence band 
maximum (VBM). The relative positions of the conduction band edges at the $\Lambda$, K and
$\Sigma$ points are also very different from the full DFT calculation. 
Similarly, in graphene, excluding $\Delta V_{xc}$ overestimates the 
ionization potential and the band splittings.
The band structure obtained from the eigenvalues obtained by diagonalizing
the constructed Hamiltonian described previously is also plotted in Fig. \ref{fig2}.
As can be seen, these band structures are in excellent agreement with the full
calculation for MoS$_2$ and graphene. A slight difference upto 5-10 meV is 
found due to a small rearrangement of charge in the bilayer or trilayer as 
compared to the sum of
the constituent layer charge densities. Hence to obtain the {\em quantitative} layer
dependence, both the effects of quantum confinement and non-linearity of the
exchange-correlation functional need to be accounted for.

The band structures of MoS$_2$ in Fig. \ref{fig2} show a transition from a direct 
band gap at K in the
monolayer to an indirect band gap from $\Gamma$ to $\Lambda$ in the bilayer.
The transition is driven by the large splitting of the VBM at the $\Gamma$ 
point and CBM at the $\Lambda$
point. The K point VBM and CBM on the other hand split only slightly. The
amount by which a band splits depends on the off-diagonal elements of the
multilayer Hamiltonian represented in the basis of the constituent layer
wavefunctions and the overlap between the wavefunctions of the constituent
layers.  The VBM and CBM of the monolayer at K are localised in space, have a
large Mo d orbital character [see Fig \ref{fig3}]. The CBM at $\Lambda$ and VBM
at $\Gamma$ on the other hand have a strong S p$_z$ character [see Fig
\ref{fig3}]. They are hence more delocalized in the out-of-plane direction and
hybridize more with the wavefunctions of other layers than the wavefunctions at
K. This leads to the large splittings in the band structure
at these points for the bilayer and trilayer [see Fig \ref{fig2} (b) and (c)]. The VBM
and CBM at K, VBM at $\Lambda$ and CBM at $\Gamma$ show little to no spliting
in the band structures owing to their localized nature [see Fig \ref{fig3}].

\section{Quasiparticle band structures}
We can extend this method to calculate the quasiparticle energies and band structures.
The DFT eigenvalues can be corrected by using the GW approximation to 
the electron self-energy,
$\Sigma$ \cite{PRB.Hybertsen,PRL.Hybertsen}. 
For the bilayer, quasiparticle eigenvalues are given by: 
\begin{equation*}
\label{eqn2}
\epsilon^{\mathrm{QP}}_i = \epsilon^{\mathrm{DFT}}_i+ \bra{\psi^{bi}_{i}}\Sigma^{bi}(\epsilon^{\mathrm{QP}}_i) - V^{bi}_{xc}\ket{\psi^{bi}_{i}}
\end{equation*}
where $\ket{\psi^{bi}_{i}}$ is the DFT
wavefunction corresponding to the eigenvalue $\epsilon_{i}^{\mathrm{DFT}}$ and 
$\epsilon^{\mathrm{QP}}_i$ is the corresponding quasiparticle energy. 
Evaluating $\Sigma^{bi}$ with the one-shot $\mathrm{G}_{0}\mathrm{W}_{0}$ method within the 
GPP approximation 
\cite{PRB.Hybertsen, PRL.Hybertsen} requires the bilayer charge density, the
bilayer irreducible polarizability, $\chi^{bi}_{0}$, and wave functions of the bilayer, $\psi^{bi}$.

We now show that all the required quantities can be approximated from 
the quantities obtained from a monolayer calculation, say on layer 'a'.
The procedure to obtain $\psi^{bi}$ and $\rho^{bi}$ is as described before.
The bilayer self-energy can be written as a sum over the individual 
self-energies of the layers and a correction term:
\begin{equation} 
  \label{eqn3}
  \bra{\psi^{bi}}\Sigma^{bi}\ket{\psi^{bi}} = \bra{\psi^{bi}}\Sigma^{a}_{\mathrm{GW}} + \Sigma^{b}_{\mathrm{GW}}\ket{\psi^{bi}} + \bra{\psi^{bi}}\Delta \Sigma\ket{\psi^{bi}}
\end{equation}
$\bra{\psi^{bi}}\Sigma^{a}_{\mathrm{GW}}\ket{\psi^{bi}}$ can be computed directly from 
monolayer irreducible polarizability, $\chi^a_{0}$, $\rho^a$ and
$\psi^a$. To compute $\bra{\psi^{bi}}\Sigma^{b}_{\mathrm{GW}}\ket{\psi^{bi}}$, 
we obtain $\chi^b_{0}$, $\rho^b$ and $\psi^b$ by applying transformations similar 
to the ones described above to $\chi^a_{0}$,
$\rho^a$ and $\psi^a$ respectively.  The correction term, 
$\bra{\psi^{bi}}\Delta\Sigma\ket{\psi^{bi}}$ contains
information on the interaction between the layers. Due to the weak coupling
between the layers, we expect $\bra{\psi^{bi}}\Delta\Sigma\ket{\psi^{bi}}$ to be a small correction (compared to $\bra{\psi^{bi}}\Sigma^{a}+\Sigma^{b}\ket{\psi^{bi}}$). We can
evaluate $\Delta\Sigma$ at various levels of approximation. It can
be evaluated at the DFT level by $\Delta\Sigma=\Delta V_{xc}$; or assuming
just exchange interaction between the layers $\Delta\Sigma_x = \Sigma^{bi}_{x}
- \Sigma^{a}_{x} - \Sigma^{b}_{x}$;
or within the static limit of GW (COHSEX) $\Delta \Sigma_{\mathrm{COHSEX}} =  
\Sigma^{bi}_{\mathrm{COHSEX}} - \Sigma^{a}_{\mathrm{COHSEX}} - 
\Sigma^{b}_{\mathrm{COHSEX}}$; or within full GW. The only additional quantity
needed in some of these approximations is, $\chi^{bi}_{0}$, which can be approximated as a sum of the irreducible polarizability 
of constituent layers:
$\chi^{bi}_{0} = \chi^a_{0} + \chi^b_{0}$ \cite{PRL.Lischner,NMat.Ugeda}. 
This method can easily be extended to calculate band structures of n layers by 
computing $\bra{\psi^{n}}\Sigma^n\ket{\psi^{n}} = \bra{\psi^{n}}\sum\limits_{i=1}
^{N}\Sigma^i_{\mathrm{GW}}\ket{\psi^{n}} + \bra{\psi^{n}}\Delta \Sigma\ket{\psi^{n}}$.
$\Delta \Sigma = \Sigma^n - \sum\limits_{i=1}^{N}\Sigma^i$ can then be evaluated 
at an appropriate level of approximation.

\begin{table}
\centering
\begin{tabular}{c@{\hskip 0.07in}r@{\hskip 0.07in}r@{\hskip 0.07in}r@{\hskip 0.07in}|r@{\hskip 0.07in}r@{\hskip 0.07in}r}
\hline
\hline
\multicolumn{1}{c}{$\Delta\Sigma$} & \multicolumn{3}{c@{\hskip 0.07in}}{Bilayer} &  \multicolumn{3}{c}{Trilayer} \\
(12$\times$12$\times$1)& \multicolumn{1}{c@{\hskip 0.07in}}{IP} & \multicolumn{1}{c@{\hskip 0.07in}}{EA} & \multicolumn{1}{c@{\hskip 0.07in}}{Gap}
& \multicolumn{1}{c@{\hskip 0.07in}}{IP} & \multicolumn{1}{c@{\hskip 0.07in}}{EA} & \multicolumn{1}{c@{\hskip 0.002in}}{Gap}
         \\
\hline
$\Delta V_{xc}$ & 5.49 & 3.38 & 2.11 & 4.84 & 2.75 & 2.09 \\
$\Delta\Sigma_{x}$ & 7.09 & 4.26 & 2.83 & 7.04 & 4.56 & 2.49 \\
$\Delta\Sigma_{\mathrm{COHSEX}}$ & 5.65 & 3.55 & 2.10 & 4.98 & 3.33 & 1.65 \\
$\Delta\Sigma_{\mathrm{GW}}^{\mathrm{800}}$ & 6.16 & 3.99 & 2.17 & 5.86 & 4.11 & 1.75 \\
Full  & 6.17 & 4.03 & 2.14 & 5.87 & 4.17 & 1.70 \\
\hline
\multicolumn{1}{c}{$\Delta\Sigma$} & \multicolumn{3}{c@{\hskip 0.07in}}{Bilayer} &  \multicolumn{3}{c}{Trilayer} \\
(24$\times$24$\times$1)& \multicolumn{1}{c@{\hskip 0.07in}}{IP} & \multicolumn{1}{c@{\hskip 0.07in}}{EA} & \multicolumn{1}{c@{\hskip 0.07in}}{Gap} 
& \multicolumn{1}{c@{\hskip 0.07in}}{IP} & \multicolumn{1}{c@{\hskip 0.07in}}{EA} & \multicolumn{1}{c@{\hskip 0.002in}}{Gap}
         \\
\hline
$\Delta\Sigma_{\mathrm{GW}}^{\mathrm{800}}$ & 6.05 & 4.03 & 2.02 & 5.81 & 4.09 & 1.72 \\

\hline
\hline
\end{tabular}
\caption{ Ionization potential (IP), electron affinity (EA) and band gap (in eV) of bilayer 
and trilayer MoS$_2$ evaluated using the constructed $\Sigma$ for various 
approximations of $\Delta\Sigma$ described in the text. The top section compares the 
results from full calculations performed with 12$\times$12$\times$1 sampling, 6000 bands to 
the results from various approximations of $\Delta\Sigma$. 
$\Delta\Sigma_{\mathrm{GW}}^{\mathrm{800}}$
 denotes $\Delta\Sigma$ evaluated at the GPP level with 800 bands. 
The bottom section are the results obtained by applying our scheme to a 
fully converged monolayer calculation.
}
\end{table}

\begin{figure}
 \centering
 \includegraphics[scale=0.43]{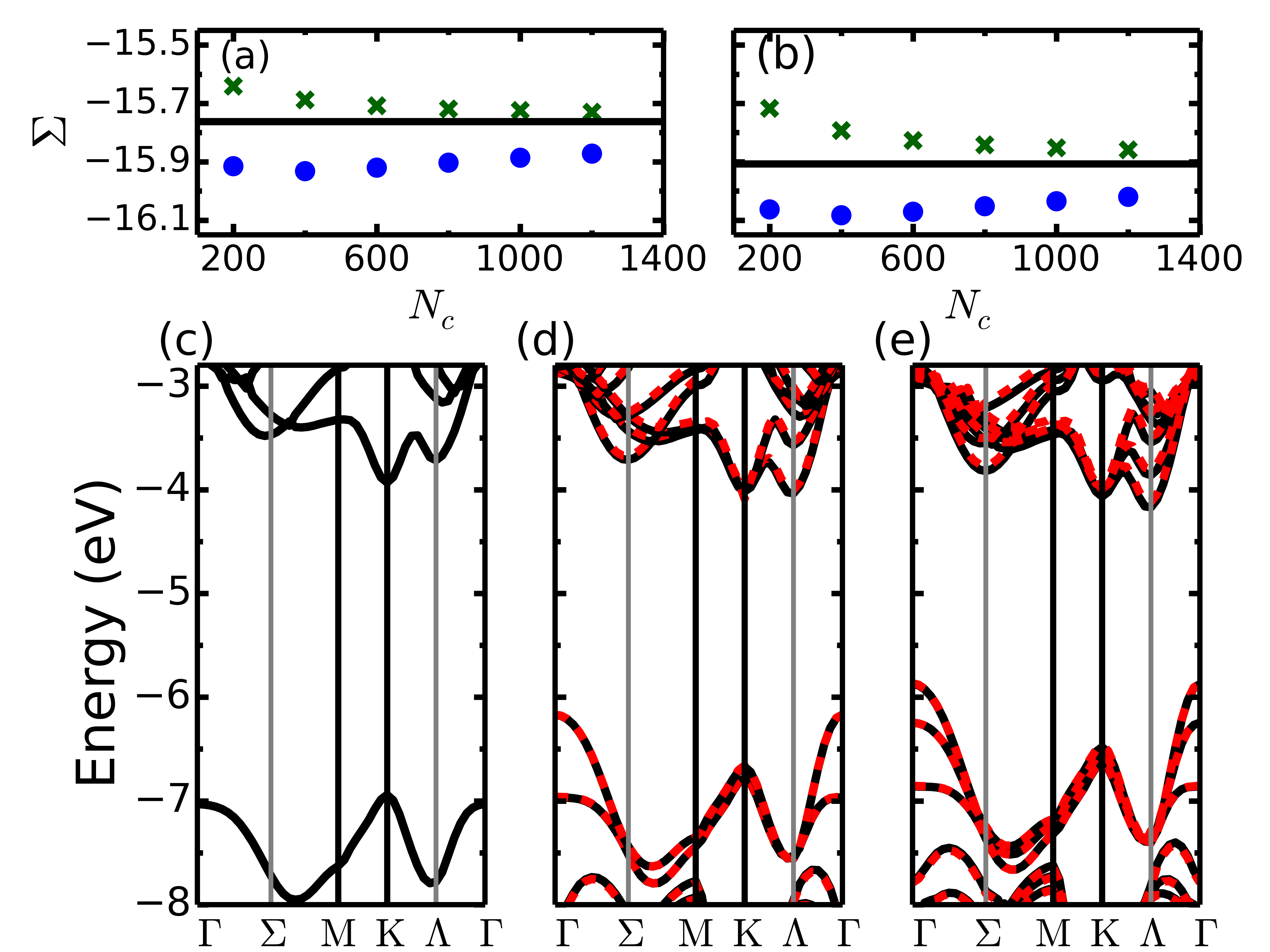}
 \caption{\label{fig4} (Color online) (a) and (b) Convergence of the constructed
 $\bra{\psi_{CBM}}\Sigma^{bi}\ket{\psi_{CBM}}$ and $\bra{\psi_{CBM}}\Sigma^{tri}\ket{\psi_{CBM}}$ with the number of bands for the CBM at
$\Lambda$ (green crosses) . The horizontal line is the converged result from full
calculations. The blue circles show the slow convergence of the full  calculation with the usual procedure.
(c) ,(d) and (e) plot the quasiparticle band structure of monolayer, bilayer
 and trilayer MoS$_2$ respectively. The black solid line indicates the result
 from separate GW calculations. The red dashed line shows the result obtained using the
 scheme described in the text. }
\end{figure}

Table 1 shows the ionization potential, electron affinity and band gap for
bilayer and trilayer MoS$_2$ for different approximations to $\Delta \Sigma$. We compare 
them to the full GW calculation on these systems. 
$\Delta \Sigma$ = $\Delta V_{xc}$ and $\Delta \Sigma
=\Delta\Sigma_{\mathrm{COHSEX}}$ show good agreement with the converged gap for
the bilayer but fail to give the right ionization potential (IP) and electron
affinity (EA). The COHSEX approximation to $\Delta \Sigma$ works well for the
band gap in trilayer too, but again falls short in the ionization potential and
electron affinity. The difference in band gap obtained using $\Delta \Sigma
=\Delta\Sigma_{\mathrm{COHSEX}}$ and $\Delta \Sigma=\Delta\Sigma_{x}$ shows that 
correlation plays an important role in the interaction between the layers.
As the next level of approximation, we compute
$\Delta\Sigma$ using GW with a few number, $N_c$, of
conduction states. 
We find that a small
$N_{c}$ = 800 is sufficient to converge the bilayer and
trilayer MoS$_2$ self-energies. Fig \ref{fig4} (a) and  \ref{fig4} (b) show this 
convergence. 
It should be noted that the small number of states are used here only to compute the 
$\Delta \Sigma$. We do not calculate $\chi^{bi}_{0}$ with a few unoccupied states.
The bilayer irreducible polarizability is constructed with 
the converged monolayer irreducible polarizabilities.
With this approximation, we obtain the IP, EA and band 
gap in good agreement with the
full calculation [see Table 1]. A major computational bottleneck in
performing separate GW calculations for the monolayer, bilayer and trilayer is
the generation of the large number of unoccupied bands on a fine k-point grid
and using these to compute the irreducible polarizability. This scheme completely does
away with the need to regenerate the unoccupied bands and the polarizability
for the bilayer and trilayer once we have the same for a monolayer. 
Fig \ref{fig4} (d) and (e) compare the bilayer and trilayer MoS$_2$ quasiparticle band 
structures obtained using this scheme with those obtained from full calculations 
on these. The eigenvalues are in good agreement with the full GW
calculation, upto 100 meV, and obtained at a small fraction of the computation
cost. Note that the results in Fig \ref{fig4} and Table 1 were obtained with slightly 
softened convergence parameters (see Computation details section). We also perform
a monolayer calculation with the fully converged parameters
and use this scheme to derive the band gap, IP and EA
for bilayer and trilayer [see Table 1]. The converged IP, EA and band gap for the monolayer 
are found to be 6.76, 4.02 and 2.74 eV respectively. The band gap for the monolayer is found 
to be in good agreement with previous calculations \cite{PRL.Qiu,PRB.Qiu}.

Table 2 compares the IP, EA and band gap for bilayer and trilayer phosphorene obtained
using this scheme to those obtained from the full calculation on these. Here we 
find that $N_c$ = 300  is sufficient to converge the bilayer and trilayer self-energies.
The COHSEX approximation to $\Delta \Sigma$ again shows good agreement with the full 
calculation for the band gap (upto 100 meV), but fails to give the right IP and EA. 
Evaluating $\Delta \Sigma$ at the COHSEX level thus seems to be a consistent approximation 
to obtain the converged quasiparticle band gap.


\begin{table}
\centering

\begin{tabular}{c@{\hskip 0.07in}r@{\hskip 0.07in}r@{\hskip 0.07in}r@{\hskip 0.07in}|r@{\hskip 0.07in}r@{\hskip 0.07in}r}
\hline
\hline
\multicolumn{1}{c}{$\Delta\Sigma$} & \multicolumn{3}{c@{\hskip 0.07in}}{Bilayer} &  \multicolumn{3}{c}{Trilayer} \\
& \multicolumn{1}{c@{\hskip 0.07in}}{IP} & \multicolumn{1}{c@{\hskip 0.07in}}{EA} & \multicolumn{1}{c@{\hskip 0.07in}}{Gap}
& \multicolumn{1}{c@{\hskip 0.07in}}{IP} & \multicolumn{1}{c@{\hskip 0.07in}}{EA} & \multicolumn{1}{c@{\hskip 0.002in}}{Gap}
         \\
\hline
$\Delta V_{xc}$ & 5.47 & 3.71 & 1.76 & 5.12 & 3.74 & 1.38 \\
$\Delta\Sigma_{x}$ & 6.45 & 4.62 & 1.83 & 6.66 & 4.86 & 1.80 \\
$\Delta\Sigma_{\mathrm{COHSEX}}$ & 5.27 & 3.69 & 1.58 & 4.69 & 3.60 & 1.09 \\
$\Delta\Sigma_{\mathrm{GW}}^{\mathrm{300}}$ & 5.73 & 4.17 & 1.56 & 5.50 & 4.25 & 1.25 \\
Full  & 5.73 & 4.20 & 1.53 & 5.49 & 4.29 & 1.20 \\
\hline
\hline
\end{tabular}
\caption{ Ionization potential (IP), electron affinity (EA) and band gap (in eV) of bilayer
and trilayer phosphorene evaluated using the constructed $\Sigma$ for various
approximations of $\Delta\Sigma$ described in the text. $\Delta\Sigma_{\mathrm{GW}}^{\mathrm{300}}$
 denotes $\Delta\Sigma$ evaluated at the GPP level with 300 bands. }
\end{table}

\section{Interlayer spacing dependence in bilayer}

\begin{figure}
 \centering
 \includegraphics[scale=0.17]{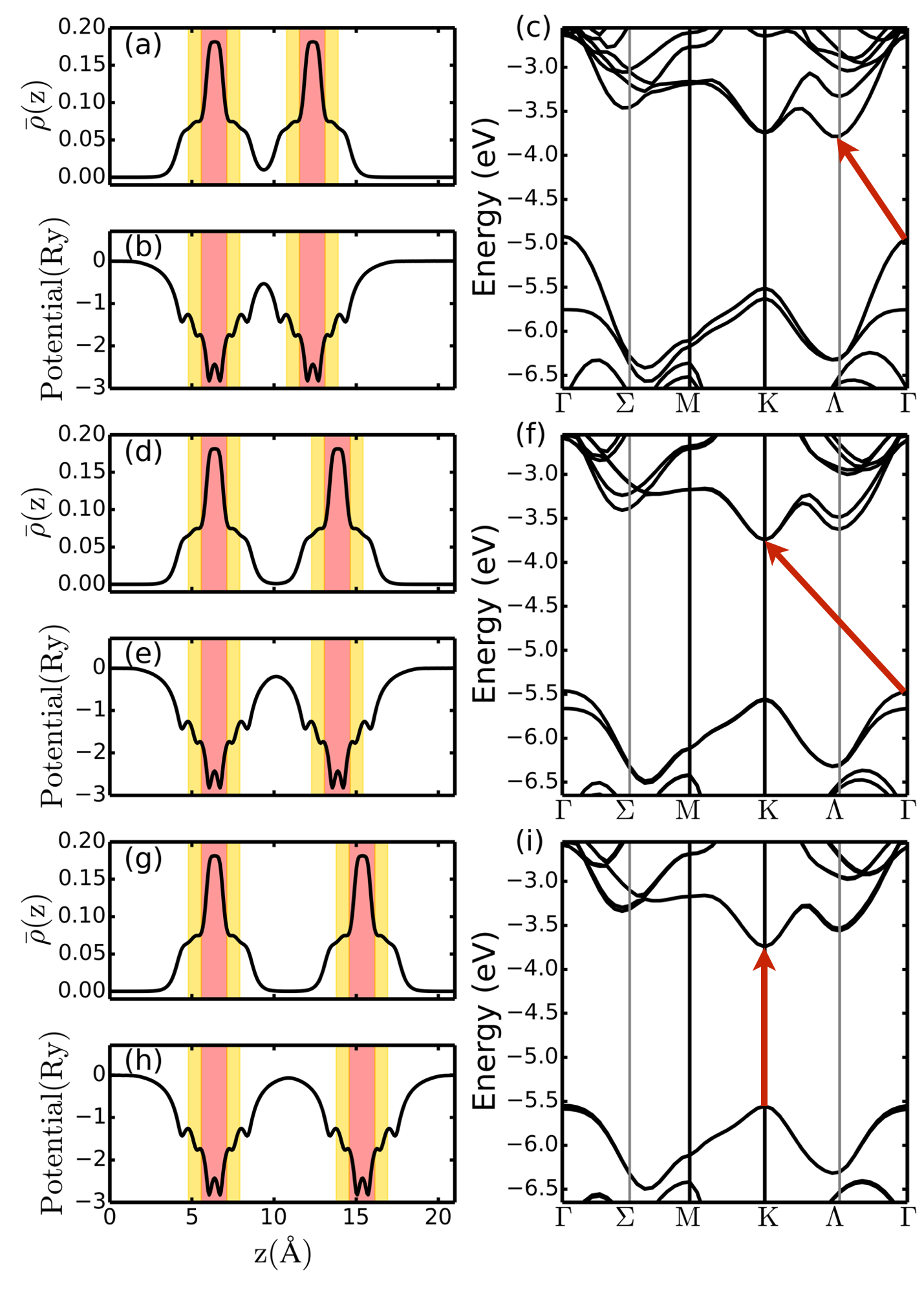}
 \caption{\label{fig5} (Color online) 
 (a), (b) and (c) DFT charge density, potential and band structure of bilayer 
 MoS$_2$ at the equilibrium interlayer spacing of 6.0 \AA, respectively.
 (d), (e) and (f) For interlayer spacing of 7.5 \AA.
 (g), (h) and (i) For interlayer spacing of 9.0 \AA.
  }
\end{figure}

\begin{figure}
 \centering
 \includegraphics[scale=0.4]{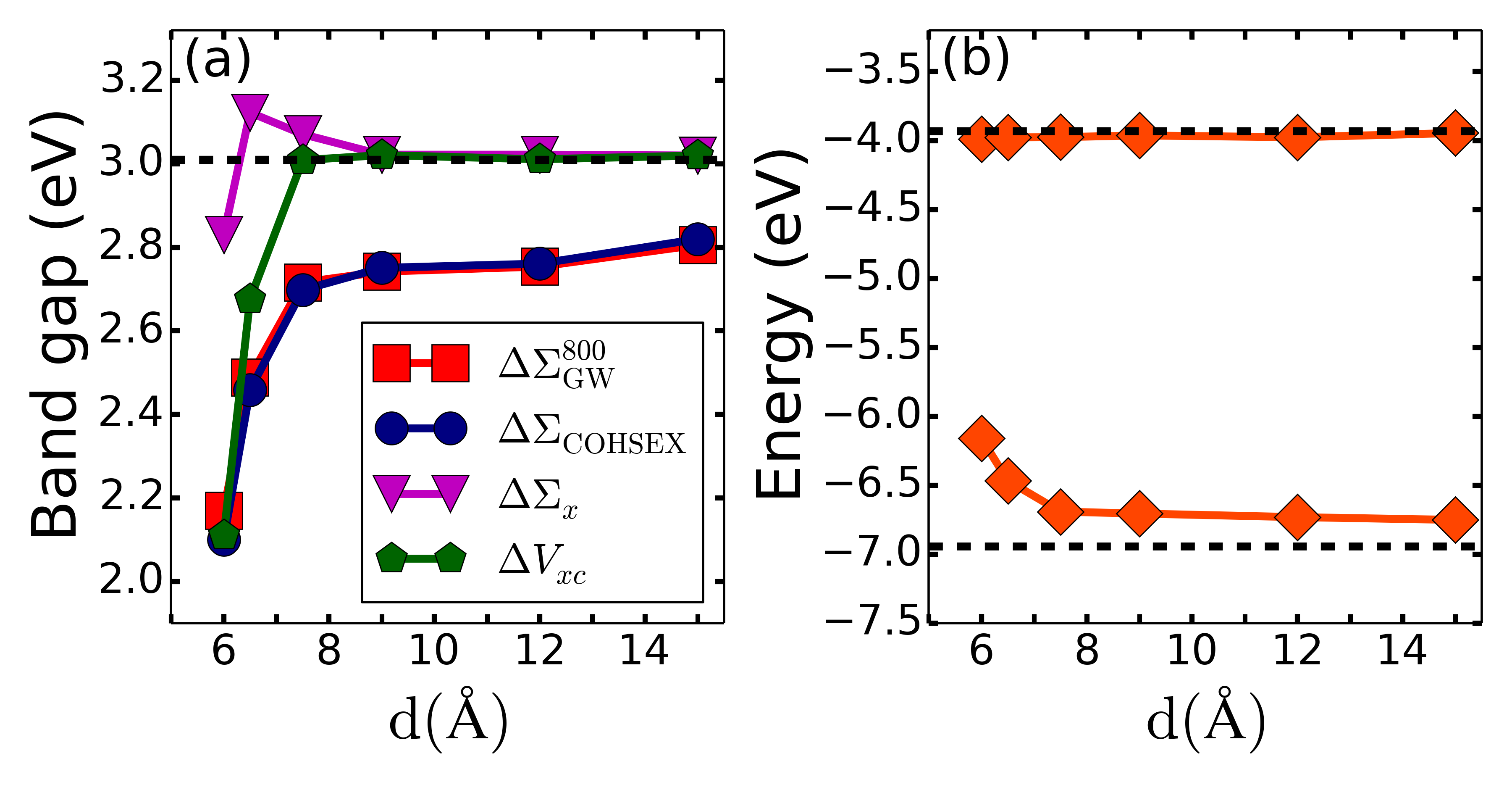}
 \caption{\label{fig6} (Color online)
 (a) Band gap of bilayer MoS$_2$ as a function of interlayer spacing evaluated using 
 the constructed $\Sigma$ for various approximations to $\Delta \Sigma$. The dashed line 
 marks the monolayer band gap within GW.
 (b) VBM and CBM of the bilayer as a function of interlayer spacing, evaluated
 using $\Delta \Sigma = \Delta \Sigma_{\mathrm{GW}}^{800}$. The dashed line marks the 
 GW VBM and CBM of a monolayer.
  }
\end{figure}

The transition in bilayer MoS$_2$, from interacting monolayers to non-interacting 
ones can be studied by gradually increasing the interlayer spacing. 
Fig. \ref{fig5} shows the evolution of the DFT band structure, charge density 
and potential of bilayer MoS$_2$ 
as a function of increasing interlayer spacing, d. The equilibrium spacing 
is $\mathrm{d}_{\mathrm{eq}} = 6$ \AA. 
As the spacing between the layers increase, the
interaction between them weakens and the splitting of the 
valence bands at $\Gamma$ and the conduction bands at $\Lambda$ reduces.
This leads to a band gap transition from $\Gamma-\Lambda$ to 
$\Gamma-\mathrm{K}$ to $\mathrm{K}-\mathrm{K}$. 
At d = 7.5 $\mathrm{\AA}$, the charge densities of the two layers stop overlapping, but 
the potential barrier between the layers is not zero. At this point the layers are weakly 
interacting and the nature of interaction within DFT
is purely due to quantum confinement; $\Delta V_{xc}$ is zero. 
At d = 9 $\mathrm{\AA}$ and above, the two layers are completely non-interacting at the DFT level 
and the gap is that of monolayer MoS$_2$. 

We construct the bilayer self energy at different interlayer spacings using the various approximations 
to $\Delta \Sigma$. The gap thus computed is shown in Fig. \ref{fig6} (a). 
In the approximation of 
$\Delta \Sigma = \Delta \Sigma_x$ and $\Delta \Sigma = \Delta V_{xc}$, the layers become non interacting once 
the charge densities of the two layers stop overlapping. Thus the gap in this approximation goes to the monolayer gap 
for spacing larger than d = 7.5 \AA. The approximations of $\Delta \Sigma = \Delta \Sigma_{\mathrm{COHSEX}}$
and $\Delta \Sigma = \Delta \Sigma_{\mathrm{GW}}^{800}$ include the long range correlation interaction between 
the layers. The band gap computed in these approximations thus show a slower convergence to the monolayer gap with 
increasing interlayer spacing. Note that these calculations are performed using a coarser 12$\times$12$\times$1 
k-point sampling and 6000 bands, leading to an overestimate of the monolayer gap in Fig. \ref{fig6}. 
Fig. \ref{fig6} (b) shows the VBM and CBM levels computed using $\Delta \Sigma = \Delta \Sigma_{\mathrm{GW}}^{800}$, 
as a function of increasing interlayer spacing.
The bilayer CBM shows a weak dependence on the interlayer spacing and is already
aligned with the monolayer CBM, while the
bilayer VBM shows a slow convergence towards the monolayer VBM as the spacing is increased. 
This is similar to the effect of a metallic substrate on a molecule, where DFT shows no renormalization 
of the band gap but accounting for correlation effects in GW shows a significant renormalization 
\cite{PRL.Thygesen,PRL.Neaton,PRL.Rohlfing}. The renormalization of the molecular levels is due to 
screening from image charge effects, arising from the metal substrate. In the bilayer MoS$_2$ system, 
similarly, one monolayer feels the screening from the other.

\section{Conclusion}
We studied the origin of layer dependence in band structures
of 2D layered materials and developed a scheme to derive multilayer 
properties from calculations on a monolayer alone. We showed that the observed trend 
in layer dependence within DFT is a combined effect of quantum confinement and non-linearity of 
the DFT exchange-correlation functional.  We
also constructed the electron self energy for multilayers in terms of monolayer 
irreducible polarizability, charge density and wavefunctions. The DFT and quasiparticle 
band structures obtained using this scheme are in excellent agreement with those 
from the full calculation. The advantage of this scheme is that it can provide accurate results
operating at a small fraction of the computation cost of a full calculation on multilayer systems.
We show that using this scheme, one is able to capture the long range correlation effects 
within GW which leads to a significant renormalization of the gap even when DFT 
finds the layers to be non-interacting at larger interlayer spacings.
This scheme can also be useful to study the variation in band structure as 
a function of stacking between the layers. 
Furthermore, it can be extended to study the nature of interaction between 
layers in heterostructures of these materials. It paves a way to derive properties of a 
heterostructure from just unit cell calculations on the constituent materials. 
This scheme is a promising tool to 
study multilayers and layer dependence of properties in the growing family of 2D layered materials.



%
\end{document}